\documentclass[%
 reprint,
nofootinbib,
 amsmath,amssymb,
 aps, prl,
]{revtex4-1}

\newcommand{\bR}{\mathbb{R}}
\newcommand{\bE}{\mathbb{E}}

\newcommand{\cN}{{\mathcal{N}}}

\newcommand{\cosnet}{Cos-net }

\newcommand{\pbi}{P({\bf b}_i)}
\newcommand{\fbi}{F({\bf b}_i)}

\newcommand{\bal}{\begin{align}}
\newcommand{\eal}{\end{align}}
\newcommand{\bea}{\begin{eqnarray}}
\newcommand{\eea}{\end{eqnarray}}

\usepackage{xcolor}

\newcommand{\be}{\begin{equation}}
\newcommand{\ee}{\end{equation}}

\newcommand{\IfN}{I^{(4)}_{N}}
\newcommand{\IfIB}{I^{(4)}_{IB}}

\usepackage{graphicx}
\usepackage{dcolumn}
\usepackage{bm}

\begin{document}

\preprint{APS/123-QED}

\title{Building Quantum Field Theories Out of Neurons}

\author{James Halverson}

\affiliation{%
\vspace{.2cm}
 $^1$The NSF AI Institute for Artificial Intelligence and Fundamental Interactions \vspace{.2cm} \\ 
 $^2$Department of Physics \\
 Northeastern University \\ 
 Boston, MA 02115
 }%


\begin{abstract}
    An approach to field theory is studied in which fields are comprised of $N$ constituent random neurons. Gaussian theories arise in the infinite-$N$ limit when neurons are independently distributed, via the Central Limit Theorem, while interactions arise due to finite-$N$ effects or non-independently distributed neurons. Euclidean-invariant ensembles of neurons are engineered, with tunable two-point function, yielding families of Euclidean-invariant field theories. Some Gaussian, Euclidean invariant theories are reflection positive, which allows for analytic continuation to a Lorentz-invariant quantum field theory. Examples are presented that yield dual theories at infinite-$N$, but have different symmetries at finite-$N$. Landscapes of classical field configurations are determined by local maxima of parameter distributions.
    Predictions arise from mixed field-neuron correlators. Near-Gaussianity is exhibited at large-$N$, potentially explaining a feature of field theories in Nature.
\end{abstract}

\maketitle


\textbf{Introduction.}--- Deep learning has succeeded in a variety of domains, such as game play and natural language processing; e.g., \cite*{Silver2016,*Silver2017} and \cite*{vaswani2017attention,*DBLP:journals/corr/abs-2005-14165}. It has been applied broadly within the physical sciences \cite{carleo2019machine}, and even in pure mathematics, including both machine-driven results (e.g., \cite*{hughes2016neural,*jejjala2019deep,*gukov2021learning}) and human-in-the-loop conjecture generation \cite{Carifio:2017bov,* Brodie:2019dfx,*Davies2021}. 

These successes are built largely on a common foundation, deep neural networks, which have received increased theoretical scrutiny in recent years in an effort to better understand empirical results. Seminal results include a correspondence  (NNGP) \cite{neal,williams,lee} between between neural networks with width $N\to\infty$ and Gaussian processes,  as well as a relationship between infinite width gradient descent and kernel methods, via the Neural Tangent Kernel \cite{NTK,lee2019wide}. The quite-general \cite{Matthews2018GaussianPB,Novak2018BayesianCN,GarrigaAlonso2019DeepCN,yangTPorig,yangTP1,yangTP2} NNGP correspondence is striking: it relates neural networks to Gaussian densities over functions, which are akin to free Euclidean theories that are not necessarily quantum. At large-but-finite $N$, these network ensembles exhibit small non-Gaussianities that motivate the use of field theory techniques, a growing area of research \cite{Dyer:2019uzd,Yaida:2019sjo,Halverson:2020trp,Bachtis:2021xoh,Maiti:2021fpy,Roberts:2021fes,Erdmenger:2021sot,Erbin:2021kqf,Grosvenor:2021eol}. In particular, though a parameter-space non-Lagrangian approach exists, actions governing network output statistics may be modeled \cite{Halverson:2020trp} and in some cases computed exactly \cite{zavatone2021exact}.  

In this Letter, we instead study a neural network approach to quantum field theory. It does not rely directly on a Lagrangian. Fields will inherit their randomness from the non-Gaussian random neurons that construct them,  and exhibit near-Gaussianity in the large-$N$ limit. Interactions arise from breaking an assumption of the Central Limit Theorem (CLT), either via $1/N$ corrections or breaking neuron independence, and symmetries arise due to transformation properties of parameter densities and measures.
The choice of neuron architecture --- the way that smaller functions are composed into neurons, and the parameters that enter into them --- allows for a significant degree of engineering that takes the place of choosing an action. Architectures engineered to satisfy the Osterwalder-Schrader axioms define Lorentz-invariant quantum field theories; simple Gaussian examples are presented, as well as cases that may be engineered, due to having a tunable $2$-pt function. Large-$N$ duality arises for some architectures. Classical field configurations and landscapes are analyzed.

\vspace{.2cm}
\textbf{Building Fields out of Neurons.}--- We study a real scalar field $\phi:\bR^d \to \bR$ built out of $N$ neurons $h_i$, as 
\begin{equation}
\phi = \sum_{i=1}^N a_i h_i(x).
\end{equation}
The neurons $h_i$, a.k.a. post-activations, are themselves random functions.
The parameters $a$ are drawn i.i.d. as
$a \sim P(a)$ and the parameters of $h$ are $\theta_h \sim P(\theta_h)$, which in general will be chosen such that $h_i$ are identically, but not necessarily independently, distributed.  The full set of parameters is 
$\theta = \{a,\theta_h\}$.

The correlation functions of the fields and neurons are
\begin{align}
G^{(2n)}(x_1,\dots,x_n)&=\bE[\phi(x_1)\dots\phi(x_n)] \\
H^{(n)}_{i_1,\dots,i_n}(x_1,\dots,x_n) &= \bE[h_{i_1}(x_1)\dots h_{i_n}(x_n)],
\end{align}
respectively.
$H_{i_1,\dots,i_n}:=H^{(n)}_{i_1,\dots,i_n}(x_1,\dots,x_n)$ is 
an abbreviation that we will use often.
They are usually computed via a Feynman path integral, depending crucially 
on an action $S[\phi]$ or $S[h]$, but in this construction we may instead compute
\begin{align}
G^{(n)}(x_1,\dots,x_n)&=\frac{1}{Z_\theta} \int d\theta \, \phi(x_1)\dots \phi(x_n) P(\theta), \\
H^{(m)}_{i_1,\dots,i_n}(x_1,\dots,x_n)&=\frac{1}{Z_{\theta_h}} \int d\theta_h \, h_{i_1}(x_1)\dots h_{i_n}(x_n) P(\theta_h)
\end{align}
where $Z_\theta = \int d\theta P(\theta)$ and  $Z_\theta = \int d\theta_h P(\theta_h)$. This is the method \cite{williams}  for computing $2$-pt functions governing NNGPs. Here, the randomness of a fixed field configuration 
$\phi$ arises from how it is constructed, rather than from the distribution $\exp[-S[\phi]]$ from which it is drawn.

To simplify the presentation in this Letter, we take $P(a)$ to have 
$\bE[a^{2n+1}]=0$ for $n\geq0$, $\bE[a^2]=\sigma_a^2/N$, and $\bE[a^4]=\gamma_a^4/N^2$,
for simplicity. Then the odd correlators vanish, $G^{(2n+1)}=0$ for $n>0$, and
the $2$-pt function is  
\begin{equation}
    G^{(2)}(x_1,x_2) = \frac{\sigma_a^2}{N} \sum_{i=1}^N H_{ii}^{(2)}(x_1,x_2) = \sigma_a^2 H_{ii}^{(2)}(x_1,x_2),
\end{equation}
with no sum on the R.H.S, which holds since the neurons are identically distributed; see Supplementary Materials.

\vspace{.1cm}
\emph{Interactions.} The leading interactions are measured by the connected $4$-pt function, 
\begin{equation}
    G^{(4)}_c(x_1,\dots,x_4)= I^{(4)}_{N}  + I^{(4)}_{IB},
    \end{equation}
    where the different $4$-point interaction terms $I^{(4)}$ are 
    \begin{align}
        \label{eqn:interaction_nosums}
        \IfN &= \frac{\gamma_a^4}{N} H_{iiii} - \frac{\sigma_a^4}{N} \Big(H_{ii}^{12}H_{ii}^{34} + H_{ii}^{13}H_{ii}^{24}  + H_{ii}^{14}H_{ii}^{23}  \Big)   \\
        \IfIB &= \sigma_a^4 (1-\delta_{ij}) \left(1-\frac{1}{N}\right) \Big(H_{iijj}-H_{ii}^{12}H_{jj}^{34}  
           \nonumber \\ &\qquad + H_{ijij}-H_{ii}^{13}H_{jj}^{24} + H_{ijji}-H_{ii}^{14}H_{jj}^{23}\Big),
\end{align}
with no sum, due to the neurons being identically distributed, and we have used our shorthand.

These interactions arise due to breaking assumptions of the Central Limit Theorem (CLT). By the CLT, if the neurons are not only identically distributed, but also independent, then we must have  $G^{(4)}_c=0$ as $N\to \infty$. We see this explicitly in the interactions, 
\be 
\lim_{N\to \infty} \IfN= 0 \qquad \qquad \text{independent:}\,\,\,\, \IfIB = 0,
\ee 
where the latter follows since, e.g., \smash{$H_{iijj} = H_{ii}^{12}H_{jj}^{34}$} for independent neurons. Finite-$N$ effects and independence breaking are necessary for non-zero $\IfN$ and $\IfIB$, respectively. Some architectures have an independence limit, 
\be 
\lim_{\epsilon \to 0} \, P(\theta_h) = \prod_{i=1}^N \, P(\theta_{h_i})
\ee 
where $\epsilon$ is a real parameter in $P(\theta_h)$ and $|\epsilon| \ll 1$ corresponds to a close-to-independent architecture with perturbative contributions to $\IfIB$. For some neuron architectures, it possible to evaluate the exact expression $G^{(4)}_c$, as we will
see in examples.

\vspace{.2cm}
\emph{Symmetries.} The symmetries of the ensemble of fields $\phi$ may be established by studying the transformation properties of correlation functions, computed in parameter space. This is the symmetry-via-duality mechanism of \cite{Maiti:2021fpy}, though here we do not study function space.
The existence of a symmetry arises due to transformation properties of the parameter densities and measures. 

Space symmetries correspond to symmetries of the neuron inputs, while internal symmetries are associated to the fields. Non-abelian internal symmetries are possible \cite{Maiti:2021fpy}, but our single real scalar $\phi$ already enjoys a 
\be
\mathbb{Z}_2 \,\,\, \text{symmetry:} \qquad \phi \to -\phi,
\ee which follows from the choice $\bE[a^{2n+1}]=0$ for $n>0$.  Our focus will be on establishing Euclidean-invariance.

\vspace{.2cm}
\textbf{Neural Network Quantum Field Theories.}---
Fields built out of neurons determine a Euclidean field theory that does not necessarily admit a continuation to a well-defined Lorentz-invariant quantum field theory. For that, its correlation functions must satisfy the Osterwalder-Schrader (OS) axioms \cite{Osterwalder:1973dx,Osterwalder:1974tc}. These include a permutation invariance that our correlation functions automatically satisfy, as well as cluster decomposition, Euclidean invariance, and reflection positivity.

We thus have a method of defining a Lorentz-invariant QFT by building fields out of neurons: choose the neuron architecture such that the correlation functions satisfy the OS-axioms. Henceforth, we call such a theory a \emph{neural network quantum field theory} (NN-QFT). In this Letter we establish a mechanism for building Euclidean invariant theories, as well as reflection positivity and cluster decomposition for Gaussian theories; the latter yield NN-QFTs. More general constructions of NN-QFTs require a systematic study of architectures that yield reflection positive correlators.

\emph{Reflection Positivity.} Unitarity and the absence of negative norm states 
in the Lorentzian theory requires reflection positivty \cite{Osterwalder:1973dx}. We focus on the Gaussian case, which requires that the $2$-pt function is reflection positive (RP).
Letting $x=(\tau_x,{\bf x})$ and similarly for $y$, RP requires \cite{Osterwalder:1973dx,glimm2012quantum}
\begin{equation}
\int d^dx d^dy \, f^*(x) f(y) \, G^{(2)}(x^\theta, y) \geq 0,
\end{equation}
for any complex function $f(x)$ with support only on $\tau>0$,
where $x^\theta = (-\tau_x,{\bf x})$ is the reflection of $x$ in 
imaginary time $\tau=i t$, which is one coordinate of the $\bR^d$ neuron input. We consider translation invariant Gaussian theories, which are determined by their power spectrum $G^{(2)}(p)$. Accordingly, we write RP in terms of the power spectrum
\begin{align}
\label{eqn:powerspec_RP}
\int \frac{d^{d-1}{\bf p}}{(2\pi)^{d-1}} \Big[\int d^dx\, e^{-i {\bf p}\cdot {\bf x}} f^*(x)\Big]\Big[\int d^dy\, e^{i {\bf p}\cdot {\bf y}} f(y)\Big] \nonumber \\
\times \rho({\bf p},\tau_x+\tau_y) \geq 0,
\end{align}
where 
\begin{equation}
    \label{eqn:rho_general}
    \rho({\bf p},\tau_x+\tau_y) = \int \frac{dp_0}{2\pi} e^{ip_0(\tau_x + \tau_y)} G^{(2)}(p).
\end{equation}
If $\rho$ factorizes as 
\be 
\label{eqn:rho_fac}
\rho = A^*({\bf p},\tau_x) A({\bf p},\tau_y) B({\bf p}),
\ee $B$ 
a positive real function, then
\eqref{eqn:powerspec_RP} has positive integrand and RP holds. For instance, 
the free scalar has $G^{(2)}(p)=(p^2+m^2)^{-1}$ and therefore 
$\rho = e^{-\mu(\tau_x+\tau_y)} / 2\mu$, $\mu=\sqrt{{\bf p}^2+m^2}$,
which has the stated form.

\vspace{.2cm}
\textbf{Euclidean Invariant Ensembles of Neurons.}--- We wish to construct Euclidean-invariant neuron ensembles. They are of interest in their own right, but also for construction NN-QFTs.
Consider a neuron of the form
\be 
h_i(x) = \sum_{j=1}^k g_{ij}(\ell_j(x)),
\label{eqn:prepend_l}
\ee
with $\ell$ the so-called input layer to the neuron, and $g$, $\ell$ having their own sets of parameters $\theta_{g}$ and $\theta_{\ell}$ with $\theta_{h} = \theta_g \cup \theta_\ell$ and $\theta_g \cap \theta_\ell = \emptyset$. We have $k\neq N$ in general, but henceforth we choose $k=N$ for simplicity. In \cite{Maiti:2021fpy} it was shown that an input layer prepended to a neural network $g$ with such parameter constraints yields a translation invariant ensemble of neurons $h$ if $\ell$ yields a translationally invariant input layer ensemble; i.e., the neuron ensemble inherits translation invariance from the input ensemble. A similar calculation demonstrates the same result for the case of Euclidean invariance. 

Hence, \eqref{eqn:prepend_l} provides a way to engineer an ensemble of neurons that is Euclidean-invariant from input layers $\ell$ with the same property. There are potentially many examples, but one such $\ell$ is given by
\begin{equation}
    \ell_i(x) = F({\bf b}_i)  \cos\Big(\sum_j b_{ij}x_j + c_i\Big)
    \end{equation}
    where $\ell: \bR^d \to \bR^N$ has 
     parameters drawn as $b_{ij}\sim P(b_{ij})$, i.i.d. $c\sim U[-\pi, \pi]$, with components $i\in \{1,\dots,N\}$, $j\in \{1,\dots,d\}$, and $F({\bf b}_i)$ is a real function of $\textbf{b}_i$, the $i^{\rm th}$ row of $b_{ij}$. $\fbi$     must be chosen in any example. It provides freedom to tune the theory.
     
To establish Euclidean-invariance of the ensemble, we must study its correlation functions.
     The correlation functions of the $\ell$ are
    \begin{align}
        L^{(n)}_{i_1\dots i_n}&(x_1,\dots,x_n) =  \bE_{b,c}[\ell_{i_1}(x_1)\dots \ell_{i_n}(x_n)] \nonumber \\ = \mathbb{E}_b&\Big[F({\bf b}_{i_1}) \dots F({\bf b}_ {i_n}) \times \nonumber \\ & \mathbb{E}_c[\cos(b_{i_1j_1} x^1_{j_1} + c_{i_1})\dots \cos(b_{i_nj_n} x^n_{j_n} + c_{i_n})]\Big],
    \end{align}
    where we have emphasized that the expectation value over $c$'s may be evaluated first and have written the $j_n^\text{th}$ component of $x_n$ as $x_{j_n}^n$.
    Invariance of the correlators $L^{(n)}$ under translations $x_j \mapsto x_j + d_j$ for an arbitrary vector $d_j \in \bR^d$ holds because 
    the $\mathbb{E}_c$ expectation value is itself translation invariant: 
    \begin{align}
        &\mathbb{E}_c[\cos(b_{i_1j_1} (x_{j_1}+d_{j_1}) + c_{i_1})\dots \cos(b_{i_nj_n} (x_{j_n}+d_{j_n})  + c_{i_n})] \nonumber \\ 
        &=
        \mathbb{E}_c[\cos(b_{i_1j_1} x_{j_1} + \tilde c_{i_1})\dots \cos(b_{i_nj_n} x_{j_n}  + \tilde c_{i_n})] \nonumber \\ 
        &= 
        \mathbb{E}_{\tilde c}[\cos(b_{i_1j_1} x_{j_1} + \tilde c_{i_1})\dots \cos(b_{i_nj_n} x_{j_n}  + \tilde c_{i_n})],
    \end{align}
    summing on $j$-indices,
    where the first equality defined $\tilde c_i = b_{ij}d_j + c_i$,
    which is constant shift of $c_i$ with respect to the $\mathbb{E}_c$ 
    expectation value. The latter integrates one period of the argument with uniformly distributed $c_i$, yielding the second equality, which 
    is invariant via renaming $\tilde c$ to $c$. 
    Similarly, invariance of the neuron correlators under 
    the $SO(d)$ transformation $x_j \to R_{jk} x_k$, $R\in SO(d)$ follows from invariance of the 
    $\bE_{b}$  expectation value under the parameter redefinition $\tilde b_{ik} = b_{ij} R_{jk}$,
    which holds when $P(b) = P({\tilde b})$ and $F(\tilde {\bf b}_i) = F({\bf b}_i)$; e.g., if $P(b) = \cN(0,\sigma_b^2/d)$ and $F({\bf b}_i) = {\bf b}_i\cdot {\bf b}_i$ or $F({\bf b}_i)=1$.
    
    The two-point function of these neurons is
    \begin{align}
    L_{ij}^{(2)}(x, y)=  \frac12 \delta_{ij}\, \bE_b[F({\bf b}_i)^2\cos(b_{ik}(x_k-y_k))]
    \end{align}
    and the associated power spectrum is
    \begin{align}
    L^{(2)}_{ij}&(p) = \frac14 \delta_{ij}\, \bE_b\Big[F({\bf b}_i)^2\Big(\delta^{(d)}(b_{ik}-p_k) + \delta^{(d)}(b_{ik}+p_k)\Big)\Big] \nonumber \\
    &= \frac{1}{4Z_{{\bf b}_i}} \delta_{ij}\, \Big[ P({\bf b}_i)F({\bf b}_i)^2\big|_{b_{ik}=p_k} + P({\bf b}_i)F({\bf b}_i)^2\big|_{b_{ik}=-p_k} \Big]
    \end{align}
    where the $\delta^{(d)}$-functions refer to the $k$-index and $Z_{{\bf b}_i} = \int d{\bf b}_i \,P({\bf b}_i)$ and $P({\bf b}_i)$ obtained in general by marginalizing $P(b_{kj})$ over $k \neq i$. Choosing $F({\bf b}_i)$ and $P({\bf b}_i)$ allows for tuning of the power spectrum.
    
     The non-zero $4$-point functions are 
    \begin{align}
    L_{iiii}&= \frac18 \,\bE_b\left[F({\bf b}_i^4) \left(\cos(b_{ij}x^{++--}_j)+ {\rm 2\,\, perms.}\right)\right] \nonumber \\
    L_{iikk}&= \frac14 \, \bE_b\left[F({\bf b}_i^2)F({\bf b}_j^2) \cos(b_{ij}(x^1_j-x^2_j)\cos(b_{ij}(x^3_j-x^4_j)\right]
    \end{align}
    and similarly for $L_{ikik}$ and $L_{ikki}$.
    We have used shorthand \smash{$L_{ijkl}:=L^{(4)}_{ijkl}(x_1,x_2,x_3,x_4)$}
    and $x^{++--}:=x_1+x_2-x_3-x_4$. Translation invariance is manifest.

\vspace{.2cm}
\textbf{Explicit Examples.}--- We wish to study a number of  Euclidean-invariant field theories, as examples. We obtain such a theory for any neuron architecture $h$ of the form \eqref{eqn:prepend_l} satisfying $\theta_h = \theta_g \cup \theta_\ell$ and $\theta_g \cap \theta_\ell = \emptyset$, but henceforth we take $g_{ij} = \delta_{ij}$ for simplicity, so that the neuron $h$ is simply $\ell$. In total, we have
\be
\label{eqn:field_example}
\phi(x) = \sum_{i=1}^N \, a_i \ell_i(x).
\ee 
The two-point function is, with no summation on $i$, 
\begin{align}
G^{(2)}(x_1,x_2) & = \sigma_a^2\, L_{ii}(x_1,x_2) \nonumber  \\
 &= \frac{\sigma_a^2}{2} \, \bE_b[F({\bf b}_i)^2\cos( \sum_{k=1}^d b_{ik}(x_k-y_k))],
\end{align}
and its power spectrum is 
\begin{align}
    \label{eqn:field_powerspectrum}
G^{(2)}(p) = \frac{\sigma_a^2 (2\pi)^d}{4Z_{{\bf b}_i}} \, \Big[ P({\bf b}_i)&F({\bf b}_i)^2\big|_{b_{ik}=p_k} + (p_k\leftrightarrow -p_k)\Big]
\end{align}
 We see that $\phi$ inherits its power spectrum from the neurons.
These theories have $4$-pt interactions determined by 
\begin{align}
    \label{eqn:examplesinteraction_nosums}
    \IfN &= \frac{\gamma_a^4}{N} L_{iiii} - \frac{\sigma_a^4}{N} \Big(L_{ii}^{12}L_{ii}^{34} + L_{ii}^{13}L_{ii}^{24}  + L_{ii}^{14}L_{ii}^{23}  \Big)   \\
    \IfIB &= \sigma_a^4 (1-\delta_{ij}) \left(1-\frac{1}{N}\right) \Big(L_{iijj}-L_{ii}^{12}L_{jj}^{34}  
       \nonumber \\ &\qquad + L_{ijij}-L_{ii}^{13}L_{jj}^{24} + L_{ijji}-L_{ii}^{14}H_{jj}^{23}\Big),
\end{align}
directly expressed in terms of the $\ell$-correlators.

With this choice $g_{ij}=\delta_{ij}$, it is interesting to compare and contrast \eqref{eqn:field_example} to the usual mode expansion of a scalar theory. Notably, 1) non-Gaussianities arise in this case and 2) the comparison is obscured in more complicated setups, e.g. if $g_{ij}$ is itself a deep neural network.

\vspace{.1cm}
\emph{Gaussian NN-QFTs.} We wish to establish the existence of Gaussian NN-QFTs, i.e. field theories built out of neurons that are \emph{quantum} field theories, due to field correlators satisfying the OS axioms.

For Gaussian theories, RP is ensured by RP of the two-point function. 
In this family of examples, the RP condition \eqref{eqn:powerspec_RP} may be expressed via the computation of $\rho$ using \eqref{eqn:field_powerspectrum},
with details depending on the choice of $\pbi$ and $\fbi$. If they are even, we have 
\begin{align}
\hspace{-.5cm}
\rho({\bf p},\tau_x+\tau_y) &= \frac{\sigma_a^2\, (2\pi)^d}{2Z_{{\bf b}_i}} \int \frac{dp_0}{2\pi} e^{ip_0(\tau_x + \tau_y)} (P({\bf b}_i)F({\bf b}_i)^2)|_{b_{ik} = p_k}. \nonumber 
\end{align}
which can be evaluated with contour integration, and checked for RP for various of $\pbi$ and $\fbi$. 

The free scalar of mass $m$ is obtained (up to normalization) by taking 
\be 
\fbi = \frac{1}{\sqrt{{\bf b}_i^2 + m^2}} \qquad \qquad \pbi = \mathcal{U}(S^d_\Lambda),
\ee 
the latter a uniform distribution on a $d$-sphere of radius $\Lambda$, with power spectrum
\be 
G^{(2)}(p) = \frac{\sigma_a^2\, (2\pi)^d}{2 \,\text{vol}(S^d_\Lambda)} \, \frac{1}{p^2+m^2}.
\ee 
$\Lambda$ plays the role of momentum space cutoff and can be taken arbitrarily large. RP for the free scalar was reviewed above. Similarly, taking $d=1$  and 
\be 
F(b_i)=1 \qquad \qquad P(b_i)=\frac{1}{b_i^2 + m^2}
\ee 
then 
\be 
G^{(2)}(p) = \sigma_a^2 \, \frac{m}{p^2+m^2},
\ee 
which is the Cauchy distribution, up to normalization, and is well-known \cite{neeb2018reflection} to give a RP kernel. In $\rho$-formulation, RP arises because 
\be 
\rho(p,\tau_x+\tau_y) = \frac{\pi m \sigma_a^2}{2} \frac{e^{-\mu (\tau_x+\tau_y)}}{\mu},
\ee 
$\mu=\sqrt{p^2+m^2}$, is of the form \eqref{eqn:rho_fac}. In this example, $b_i$ and $p$ are scalars since $d=1$.

Generally, the tunable power spectrum (by choice of $\pbi$ and $\fbi$) offers freedom to engineer RP theories.

\vspace{.1cm}
\emph{Large-$N$ Duality.} Since the choice of architecture defines the field 
theory, it is natural to wonder if there are dual architectures that look different,
but nevertheless describe the same theory. To that end, we note that 
by taking $P(a) = \cN(0,\sigma_a^2/N)$, $P(b) = \cN(0,\sigma_b^2/d)$, and $F=1$, the power spectrum is 
\begin{equation}
G^{(2)}(p) = \frac{\sigma_a^2}{2Z_b} e^{-\frac12 \frac{d}{\sigma_b^2} p^2}.
\end{equation}
We call this model Cos-net, since the action function is a cosine.
The power spectrum is equivalent to that of Gauss-net \cite{Halverson:2020trp,Erbin:2021kqf}, which instead uses 
an exponential activation. In the 
large-$N$ limit , the field theories are GPs and are 
determined by their power spectrum. They are therefore dual theories, despite different constructions.

This is large-$N$ duality. Finite-$N$ effects introduce non-Gaussianities that gives rise to different correlation functions, with different symmetries. Cos-net is Euclidean-invariant in all correlators at all $N$, by construction, while Gauss-net
breaks translation invariance already in the connected four-point function, which is leading order in $1/N$. See the Supplementary Materials for details.

\vspace{.2cm}
\textbf{Classical Field Configurations, Landscapes, and Symmetry Breaking}---
Since we do not know the action, a priori, we must obtain a different way of analyzing classical field configurations. We can think of them as local maxima in probability, described not in the usual way of maximizing $\exp(-S[\phi])$, but instead in parameter space.
Consider a field constructed out of neurons, represented as 
$\phi_\theta(x), \theta \sim P(\theta),$
where our notation emphasizes that $\phi$ depends on parameters $\theta$ drawn from a distribution
$P(\theta)$. Then the set of classical configurations are of the form
\begin{equation}
S_c = \{\phi_{\theta^*} \, | \, \theta^* \in {\rm max}(P(\theta))\},
\end{equation}
where ${\rm max}(P(\theta))$ is the set of local maxima of $P(\theta)$. The dominant 
configurations (global minima) in $S_c$ are
\begin{equation}
\phi_{\theta^*_d} \qquad \qquad \theta^*_d = \underset{\theta}{\operatorname{arg\,max}} \,P(\theta).
\end{equation}
As a trivial example, an ensemble of lines $\phi(x) = \theta x$, $\theta \sim \mathcal{N}(1,1)$, has 
one classical configuration, $\phi(x) = x$.

The neurons $h$ have their own classical configurations that influence the classical configurations of the fields. Writing
\begin{equation}
\phi(x) = \sum_{i=1}^N a_i h_i(x), \qquad a \sim P(a),
\end{equation}
the coefficients $a_i$ are drawn together from the distribution $P(a)$,
and let the parameters of $h$ be $\theta_h$. Then 
\begin{equation}
S_c^h = \{h_{\theta^*_h} \, | \, \theta^*_h \in {\rm max}(P(\theta_h))\}
\end{equation}
are the classical configurations of the neuron ensemble, and we may rewrite the
classical field configurations in terms of classical neuron configurations as 
\begin{equation}
    S_c = \{\phi_{{a^*},h_i} \, | \, a^* \in {\rm max}(P(a)), h_i \in S^h_c\}.
\end{equation}
The number of such configurations is bounded by 
\begin{equation}
|S_c| \leq |{\rm max}(P(a))| \, |S_c^h|^N,
\end{equation}
since a priori one may choose the $N$ classical neurons and maxima independently, and $P(a)$ itself has some maxima. In the case that $a_i$ are drawn i.i.d., $P(a)=\prod_i P(a_i)$ and $|{\rm max}(P(a))| = |{\rm max}(P(a_i))|^N$.

There are a number important considerations that may arise in some examples.
First, $|S_c|$ 
may in fact be much smaller than the bound due to redundancies, i.e., where the same classical field configuration admits many 
different constructions out of constituent neurons. Second, if ${\rm max}(P_a)$ is only the 
single point given by $a_i = 0\,\, \forall i$, then there is no non-trivial classical field configuration.
This arises, for instance, if the $a_i$ are drawn i.i.d. from $P(a_i)$ with a local maximum only at $a_i=0$; one such common case for deep neural networks is  $P(a_i) = \mathcal{N}(0,\sigma^2)$. It also arises for \cosnet itself, which therefore
has only a trivial classical configuration, despite having non-trivial classical configurations for its constituent neurons. Modifications of \cosnet exist with the same symmetries, but non-trivial classical configurations, by choosing $P(a_i)$ to be an even multimodal distribution.

It is also easy to arrive at landscapes of classical configurations, however. We will show this by putting a lower bound on
$|S_c|$. Consider any case where $|S_c^h|>N$ and construct $\phi$ out of a set of classical neurons ${h_i}$ without 
repeats, with i.i.d. $a_i \sim P(a)$. For $\phi$ to be classical, each $a_i$ must be a local maximum of $P(a)$,
of which there are $|{\rm max}(P(a))|$. Using this construction, we have 
\begin{equation}
|S_c| \geq |{\rm max}(P(a))|^N\, \frac{|S_c^h|!}{N!(N-|S_c^h|)!},
\end{equation}
i.e., if $P(a)$ is multimodal then the number of classical configurations is 
exponentially large in $N$. This is a prescription for constructing 
theories with large landscapes.

\vspace{.1cm}
\emph{Spontaneous Symmetry Breaking.} Consider an architecture that ensures a symmetric ensemble of fields, as measured by invariant correlation functions. Explicit symmetry breaking arises by deforming the architecture, most easily its parameter distributions, such that the correlation functions are no longer invariant; e.g., a theory with $a \sim \cN(0,\sigma_a^2/N)$ exhibits $\phi \to -\phi$ symmetry, but deforming to $a \sim \cN(\mu,\sigma_a^2/N)$ breaks it, since the deformation yields non-trivial odd-point functions.

Spontaneous symmetry breaking (SSB), on the other hand, arises when an architecture exhibits a symmetry in its correlators, but not in its classical configurations. For instance, 
a $D$-dimensional real scalar 
\be 
\phi_i(x) = \sum_{j=1}^N a_{ij}h_j(x) \qquad a \sim \cN(0,\sigma_a^2/N)
\ee 
exhibits $SO(D)$ invariance \cite{Maiti:2021fpy} due to $SO(D)$ invariance of the Gaussian. In such a case each ${\bf a}_j$ is distributed as 
\be
{\bf a}_j \sim P({\bf a}_j) \propto e^{-\frac{1}{2\sigma_a^2}{\bf a}_j \cdot {\bf a}_j},
\ee
which may be factorized into products of Gaussians for individual components. Here ${\bf a}_j$ is the $j^\text{th}$ row of the matrix $a_{ij}$.
If instead we take 
\begin{align}
\label{eqn:SSB}
{\bf a}_0 \sim P({\bf a}_0) \propto e^{-\frac{1}{2\sigma_a^2}({\bf a}_0 \cdot {\bf a}_0 - v^2)^2},
\end{align}
leaving other distributions the same,
then the correlation functions are still $SO(D)$-invariant, due to invariance of $P({\bf a}_0)$. However, the wine-bottle potential on ${\bf a}_0$  forces the modes of the distribution for ${\bf a}_0$ to live on $S^{D-1}_v$, the $(D-1)$-sphere of radius $v$. Correlation functions of fluctuations around the associated classical configurations break symmetry, corresponding to SSB. It is also interesting to turn on the breaking for multiple  ${\bf a}_j$, but this also yields subtleties that we leave for future work.

\vspace{.2cm}
\textbf{Discussion.}---
We have studied an approach to field theory in which fields are built out of neurons, which themselves are random functions that generally have non-Gaussian statistics. The neurons may themselves be deep neural networks, or simple functions, such as the ones that we have studied. Parameters appear in both the fields and neurons, and correlation functions may be computed in parameter space. Euclidean invariant ensembles of neurons were constructed, which may be used to construct Euclidean-invariant field theories in Euclidean space. Classical field configurations were studied, and a large-$N$ duality was presented. The framework may be studied even away from perturbative large-$N$ limits. 

If the Euclidean field theories constructed in this way have correlation functions that satisfy the Osterwalder-Schrader axioms, they define a neural network quantum field theory; Gaussian examples were presented. Systematically engineering NN architectures that satisfy the OS axioms, especially reflection positivity, is an interesting direction for future research. 

Field theories built out of a large number of independent neurons due make a universal prediction: they are near Gaussian fixed points, due to the Central Limit Theorem. 
If this is the origin of a near-Gaussian theory appearing in Nature, such as the Standard Model, new predictions could arise from mixed field-neuron correlators, which will be studied in future work.

\vspace{.2cm}
\noindent {\bf Acknowledgements.} We thank Sergei Gukov, Matt Schwartz, and especially Anindita Maiti and  Keegan Stoner for collaboration on related topics. We thank Tom Hartman, Austin Joyce, and Fabian Ruehle for conversations. This work is supported by NSF CAREER grant PHY-1848089 and by the National Science Foundation under Cooperative Agreement PHY-2019786 (The
NSF AI Institute for Artificial Intelligence and Fundamental Interactions).

\bibliography{refs}

\clearpage
\onecolumngrid
\renewcommand\thefigure{S\arabic{figure}}  
\renewcommand\thetable{S\arabic{table}}  
\renewcommand{\theequation}{S\arabic{equation}}
\renewcommand{\thepage}{P\arabic{page}} 
\setcounter{page}{1}
\setcounter{figure}{0}  
\setcounter{table}{0}
\setcounter{equation}{0}

\def\beq{\begin{equation}}
\def\eeq{\end{equation}}
\appendix
\section{\label{app} \large{Supplementary Material for Building Quantum Fields out of Neurons}}

\appendix

\section{Statistics of Neuron Parameters and Correlators}
In this work we consider identically distributed neurons. 
The full set of parameters $h_i$ were denoted
$\theta_h$, but for thoroughness we break these up as 
\be 
\theta_h = \{\theta_{h_1},\dots,  \theta_{h_N}\}, \qquad \qquad \theta_{h_i} = \{ \theta_{h_{i1}},\dots, \theta_{h_{ik}} \},
\ee 
where the latter are sets of parameters of individual neurons, that together comprise the entire set $\theta_h$. The partition function is 
\be 
Z_{\theta_h}=\int d\theta_h P(\theta_h) = \int \prod_{i=1}^N \prod_{j=1}^k ~d\theta_{ij} P(\theta_h),
\ee 
where $P(\theta_h)$ is a joint distribution on the $\theta_{ij}$. The marginal distributions for $\theta_{h_i}$ and $\theta_{ab}$ are given by
\be
P(\theta_{h_a}) = \int \prod_{\substack{i,j\\i \neq a}} ~d\theta_{ij} \,P(\theta_h) 
\qquad \qquad \qquad P(\theta_{ab}) = \int \prod_{\substack{i,j\\(i,j)\neq(a,b)}} ~d\theta_{ij} \,P(\theta_h).
\ee 
and associated partition functions $Z_{\theta_{h_a}} = \int \prod_{j=1}^k ~d\theta_{aj} P(\theta_{h_a})$ and $Z_{\theta_{ab}} = \int d{\theta_{ab}} \, P(\theta_{ab})$. The neurons are identically distributed if 
\begin{equation}
P(\theta_{h_a}) = P(\theta_{h_b}) \qquad \forall a,b,
\end{equation}
and they are independently distributed if 
\be
P(\theta_h) = \prod_{i=1}^N P(\theta_{h_n}).
\ee 
They are independently and identically distributed (i.i.d.) if both properties hold.

These properties have implications for correlation functions. The diagonal part of the neuron two-point function is given by 
\be 
H_{ii}^{(2)}(x_1,x_2) = \frac{1}{Z_{\theta_h}} \int \prod_{j=1}^N d\theta_{h_j} \, h_i(x_1)h_i(x_2) \, P(\theta_h) 
= \frac{1}{Z_{\theta_{h_i}}} \int d\theta_{h_i} \, h_i(x_1)h_i(x_2) \, P(\theta_{h_i}),
\ee 
where we have marginalized over the parameters not in $h_i$. For neurons that are identically distributed, the marginals $P(\theta_{h_i})$ are equal for all values of $i$, and therefore 
\be 
H_{ii}^{(2)}(x_1,x_2) = H_{jj}^{(2)}(x_1,x_2)\qquad \forall i, j.
\ee 
For identically distributed neurons, a similar calculation for the four-point function demonstrates that 
\be 
H_{iijj}^{(4)}(x_1,x_2,x_3,x_4) = H_{kkll}^{(4)}(x_1,x_2,x_3,x_4) \qquad \forall i,j,k,l.
\ee 
On the other hand, if the neurons are independent, then the correlation functions factorize, e.g.
\be 
H_{iikk}^{(4)}(x_1,x_2,x_3,x_4) =  \frac{1}{Z_{\theta_h}} \int \prod_{j=1}^N d\theta_{h_j} \, h_i(x_1)h_i(x_2) h_k(x_3)h_k(x_4)  \, P(\theta_{h_j})  = H_{ii}^{(2)}(x_1,x_2)H_{kk}^{(2)}(x_3,x_4),
\ee 
where the last identity holds because the integrals factorize.
Identities such as these are used in the main text.

\section{Large-$N$ Duality: Cos-net and Gauss-net}
\label{app:largeN}

In the text, we claimed the existence of two architectures that yield dual theories (equivalent Gaussian processes) at infinite-$N$, but have different statistics and even different symmetries at finite $N$.

The architectures are as follows. First, we have Cos-net, defined to be \eqref{eqn:field_example} with $P(b) = \cN(0,\sigma_b^2/d)$ and $F=1$, i.e.,
\be 
\phi(x) = \sum_{i=1}^N a_i l_i(x), \qquad \ell_i(x) = \sum_{j} \cos(b_{ij} x_j + c_i), \qquad \qquad a \sim \cN(0,\frac{\sigma_a^2}{N}), \,\,\, b \sim \cN(0,\frac{\sigma_b^2}{d}), \,\,\, c\sim \mathcal{U}[-\pi, \pi]. 
\ee 
The two-point function and power spectrum are given by 
\be 
\label{eqn:gausscos2pts}
G^{(2)}(x_1,x_2) = \frac{\sigma_a^2}{2} e^{-\frac{\sigma_b^2}{2d} (x_1-x_2)^2} \,  \qquad \qquad G^{(2)}(p) = \sqrt{\frac{\pi d}{2}}\frac{\sigma_a^2}{\sigma_b}e^{-\frac{d}{2\sigma_b^2}p^2}.
\ee 
On the other hand, Gauss-net with output bias turned off is defined by \cite{Halverson:2020trp}
\begin{align}
\phi(x) = &\frac{\sum_{i=1}^N a_i g_i(x)}{\sqrt{\exp[2(\sigma_c^2 + \sigma_b^2 x^2 / d)]}} \qquad \qquad g_i(x) = \sum_j \exp(b_{ij} x_j + c) \\ &a\sim \cN(0,\frac{\sigma_a^2}{2N})\,\,\, b\sim \cN(0,\frac{\sigma_b^2}{d}) \,\,\, c\sim \cN(0,\sigma_c^2),
\end{align}
which also has a two-point function and power spectrum given by \eqref{eqn:gausscos2pts}. As $N\to \infty$, these architectures yield the same Gaussian theory (which also \emph{happen} to have Gaussian $2$-pt functions), and therefore define dual theories.

Since these architectures are both built out of independent neurons, interactions in $G^{(4)}_c$ arise only from $1/N$ corrections. 
The connected correlator for Gauss-net is 
\begin{align}
    G^{(4)}|_{c, Gauss} =&\frac{1}{4 N} \sigma_a^4 \bigg[ 3e^{4\sigma_c^2} e^{-\frac{\sigma_b^2}{2d}[x^2_1 + x_2^2 + x_3^2 + x_4^2 -2x_1x_2 - 2x_1x_3 - 2x_1x_4 - 2x_2x_3 - 2x_2x_4 - 2x_3x_4 ] } \nonumber \\ &- e^{-\frac{1}{2d} \sigma_b^2
    \left((x_{1}-x_{4})^2+(x_{2}-x_{3})^2\right)}- e^{-\frac{1}{2d} \sigma_b^2
    \left((x_{1}-x_{3})^2+(x_{2}-x_{4})^2\right)}- e^{-\frac{1}{2d} \sigma_b^2
    \left((x_{1}-x_{2})^2+(x_{3}-x_{4})^2\right)} \bigg]
\end{align}
and for Cos-net it is
\begin{align}
    G^{(4)}|_{c} = &\frac{1}{8 N} \sigma_a^4 \bigg[ 3  \left(e^{-\frac{1}{2d} \sigma_b^2
   (x_{1}+x_{2}-x_{3}-x_{4})^2}+e^{-\frac{1}{2d} \sigma_b^2 (x_{1}-x_{2}+x_{3}-x_{4})^2}+e^{-\frac{1}{2d}
   \sigma_b^2 (x_{1}-x_{2}-x_{3}+x_{4})^2}\right) \nonumber \\ &-2 e^{-\frac{1}{2d} \sigma_b^2
   \left((x_{1}-x_{4})^2+(x_{2}-x_{3})^2\right)}-2 e^{-\frac{1}{2d} \sigma_b^2
   \left((x_{1}-x_{3})^2+(x_{2}-x_{4})^2\right)}-2 e^{-\frac{1}{2d} \sigma_b^2
   \left((x_{1}-x_{2})^2+(x_{3}-x_{4})^2\right)} \bigg],
\end{align}
which for Cos-net may be derived from the more general expressions in the text. We see that $G^{(4)}_c$ for Cos-net is built out of translationally invariant combinations of inputs, as it must be since it is constructed out of a Euclidean-invariant ensemble of neurons. However, the interactions for Gauss-net are clearly not translation invariant.

Thus, the duality only exists at infinite-$N$, where the architectures defined Euclidean-invariant Gaussian theories with Gaussian $2$-pt functions. Such theories do not satisfy RP, but are interesting statistical field theories in their own right, and are in fact regularly used for Bayesian inference with Gaussian processes \cite{books/lib/RasmussenW06}. Furthermore, at finite-$N$ not only do their statistics disagree, but these theories actually have different symmetries.

\end{document}